\documentclass[11pt,letterpaper]{article}
\usepackage[utf8]{inputenc}
\usepackage[english]{babel}
\usepackage{amsmath}
\usepackage{amsfonts}
\usepackage{amssymb}
\usepackage{graphicx}
\usepackage[left=3cm,right=3cm,top=2cm,bottom=3cm]{geometry}
\author{Jorge Pinochet}
\title{\textbf{Einstein ring: Weighing a star with light}}
\begin{document}

\author{Jorge Pinochet$^{1}$, Michael Van Sint Jan$^{2}$\\ \\
 \small{$^{1}$\textit{Facultad de Educación}}\\
 \small{\textit{Universidad Alberto Hurtado, Erasmo Escala 1825, Santiago, Chile.} japinochet@gmail.com}\\     	 
 \small{$^{2}$\textit{Woodtech S.A.}}\\
 \small{\textit{El Golf 150, Las Condes, Santiago, Chile}}\\}

\date{\small \today}
\maketitle

\begin{abstract}
\noindent In 1936, Albert Einstein wrote a brief article where he suggested the possibility that a massive object acted as a lens, amplifying the brightness of a star. As time went by, this phenomenon –known as gravitational lensing– has become a powerful research tool in astrophysics. The simplest and symmetrical expression of a gravitational lens is known as Einstein ring. This model has recently allowed the measurement of the mass of a star, the white dwarf Stein 2051 B. The purpose of this work is to show an accessible and up-to-date introduction to the effect of gravitational lensing, focused on the Einstein ring and the measurement of the mass of Stein 2051 B. The intended audience of this article are non-graduate students of physics and similar fields of study, and requires only basic knowledge of classical physics, modern physics, algebra and trigonometry. \\ \\

\noindent \textbf{Keywords}: General relativity, gravitational lens, Einstein ring, star’s mass measurement, undergraduate physics students. 

\end{abstract}

\maketitle

\section{Introduction}

In 1915, Albert Einstein presented a new theory of gravity, known as general relativity, which perfects and widens Isaacs Newton’s law of universal gravitation [1]. Using this theory, Einstein predicted that the trajectory of a beam of light that goes through the gravitational field of a massive body will deviate in proportion to the mass of the body. It was this prediction that allowed the first empirical confirmation of general relativity, and that made Einstein an international celebrity. This confirmation constitutes one of the greatest milestones of the 20th century. It was carries out in 1919 by an English expedition through the observation of a total solar eclipse that occurred on Prince island, off the coast of Africa [2].\\

Two decades after, in 1936, Einstein wrote a short article where he extended his prediction of 1916. In this article, he suggested that by deviating light, a massive body can act as a lens, greatly increasing a star’s brightness [3]. Although he expressed his skepticism about the possibility to directly observe this phenomenon, the great physicist underestimated the progress of astronomy in the next decades. Indeed, the gravitational lens effect, as is now known, has become a powerful tool in astronomical investigation, and has allowed the study of such diverse objects as extrasolar planets, dark matter and quasars. The last astronomical milestone based on the gravitational lens effect occurred only months ago and consisted in the first accurate measurement of the mass of a near star [4]. We are talking about Stein 2051 B, a white dwarf\footnote{A White dwarf is a cold and stable star that has exhausted its nuclear fuel and that maintains its state of equilibrium because of the electrons repelling each other by the principle of exclusion.} some 18 light-years away from us, that is part of a binary system along with a red dwarf (Stein 2051 A).\\

As is the case with many of Einstein’s great ideas, the gravitational lens effect has become a subject in the leading edge of modern physics and astronomy, that is powering a wide research program with unforeseen reach.  Because gravitational light deviation is a prediction of general relativity -a complex theory that is usually studied at the PhD level in physics or astronomy- the analysis of this effect and its important applications in astronomy are normally left out of undergraduate courses of physics and similar careers. This fact can be verified by reviewing the most widely known modern physics textbooks, and noting that the subject is absent or, at best, treated qualitatively and superficially  [5-9]. 
The goal of this article is to present an accessible and updated introduction to the gravitational lens effect. However, given the complexity and broadness of the subject, the article focuses in a particular phenomenon that allowed the measurement of the mass of the white dwarf Stein 2051 B. This phenomenon is known as Einstein ring, and is the simplest and symmetrical expression of the gravitational lens effect. To achieve the abovementioned goal, the article starts by introducing some basic notions of general relativity, presenting a heuristic derivation of Einstein’s equation for the light deviation angle, based on the principle of equivalence. Then we present a simplified analysis of Einstein’s ring and explain how it is used to determine the mass of a star. For this we examine the specific case of Stein 2051 B, using the original data taken by the team that measured the star’s mass. Finally, we briefly discuss the importance of the gravitational lens effect in astronomy and the meaning and implications of using Einstein ring to measure the mass of Stein 2051 B.\\

To make the article accessible to the broadest possible audience, we have tried to keep the technical terms to a minimum, and we have put special emphasis in mathematical simplicity. In fact, the article only assumes general knowledge of classical and modern physics and basic knowledge of algebra, geometry and trigonometry. From this perspective, we hope that the article can be rewarding for non-graduate students of physics and for non-specialists. The article may be particularly useful for teachers giving courses of astronomy or modern physics at under-graduate level, and who are looking for updated material that examines the latest advances in physics and astronomy.

\section{General relativity, the equivalence principle and the bending of light}

The so-called equivalency principle, first enounced by Einstein in 1911, and considered by him as the foundation of his theory of general relativity, states that: a homogeneous gravitational field is completely equivalent to a uniformly accelerated frame of reference. An alternative way to formulate this principle is as follows: experiments made in a frame of reference uniformly accelerating wit acceleration $\vec{a}$ with respect to an inertial frame, produces the exact same experimental results as an inertial frame of reference in a uniform gravitational field $-\vec{a}$ [8, 9]. \\

A simple example allows us to better understand the meaning of the equivalency principle. Imagine a spaceship moving with constant acceleration $g$, in a region of space away from any gravitational influence (see figure 1). Inside the ship an astronaut of mass $m$ is standing over a weighing scale. What value does the instrument show? As the only force acting upon the astronaut is the normal $N$ exerted by the scale in the same direction as $g$, from Newton’s second law we know that the value shown is $N = mg$. On the other hand, if we consider the same spaceship standing still on the surface of a planet where the gravitational acceleration is $-g$, the scale will also show the value $mg$.\\

\begin{figure}
  \centering
    \includegraphics[width=0.7\textwidth]{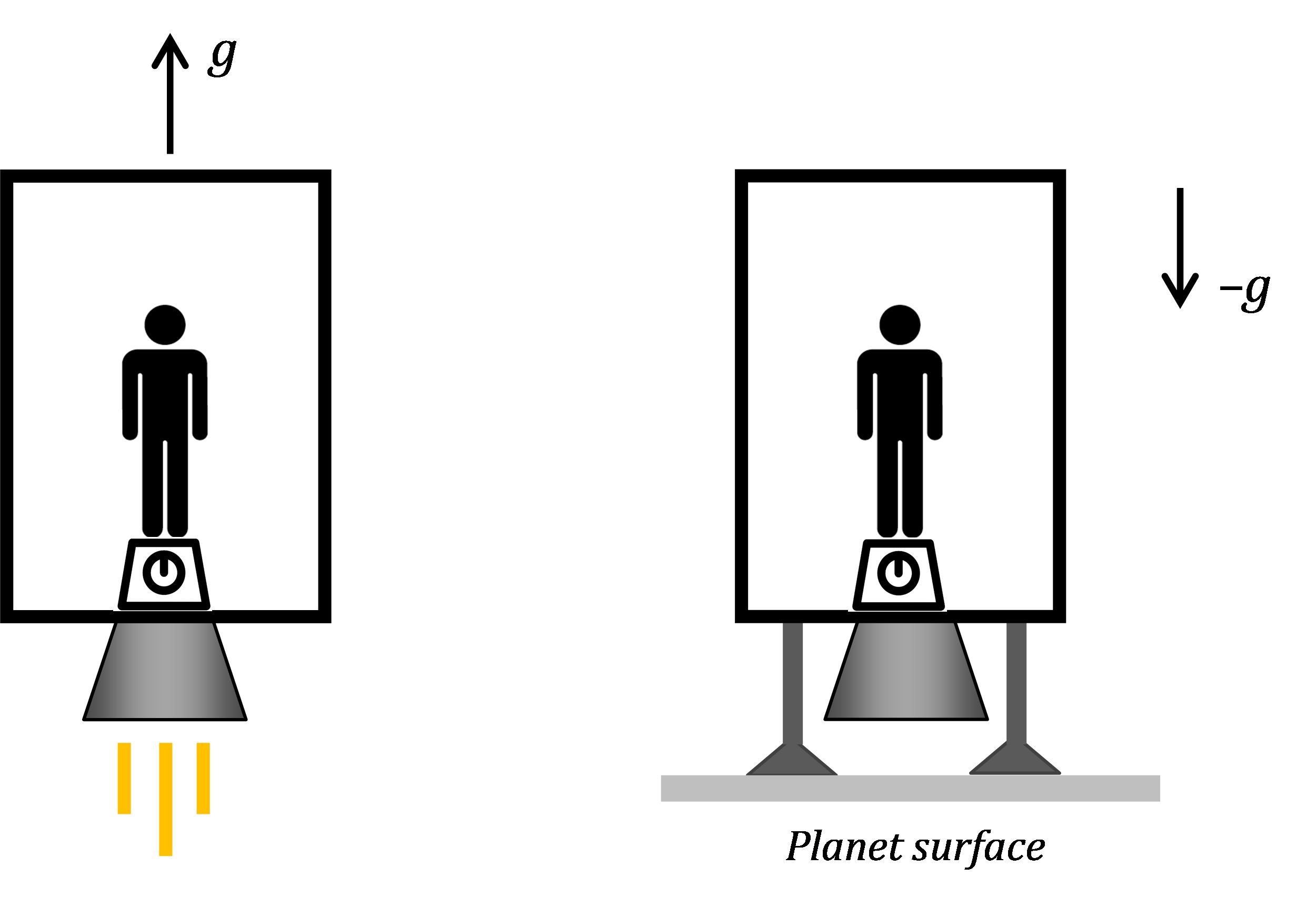}
  \caption{An astronaut is standing over a weighing scale inside a spaceship. On the left, the spaceship is travelling in space with acceleration $g$, far from any gravitational influence. On the right, the spaceship is standing on the surface of a planet with a surface acceleration of gravity $-g$. In both cases the scale shows $mg$.}
\end{figure}

From the equivalency principle, Einstein predicted that a beam of light in a gravitational field will deviate from its trajectory [10]. To see how one could arrive at this conclusion, let’s imagine that in the same spaceship from the previous example, a photon enters through a window located in the left wall, traveling a distance $x$ before arriving at the opposite wall. Figure 2 shows two possible perspectives of this photon (drawn as a red dot) in four equidistant instants in time, while the spaceship travels upward with constant acceleration of magnitude $g$.\\

\begin{figure}
  \centering
    \includegraphics[width=0.9\textwidth]{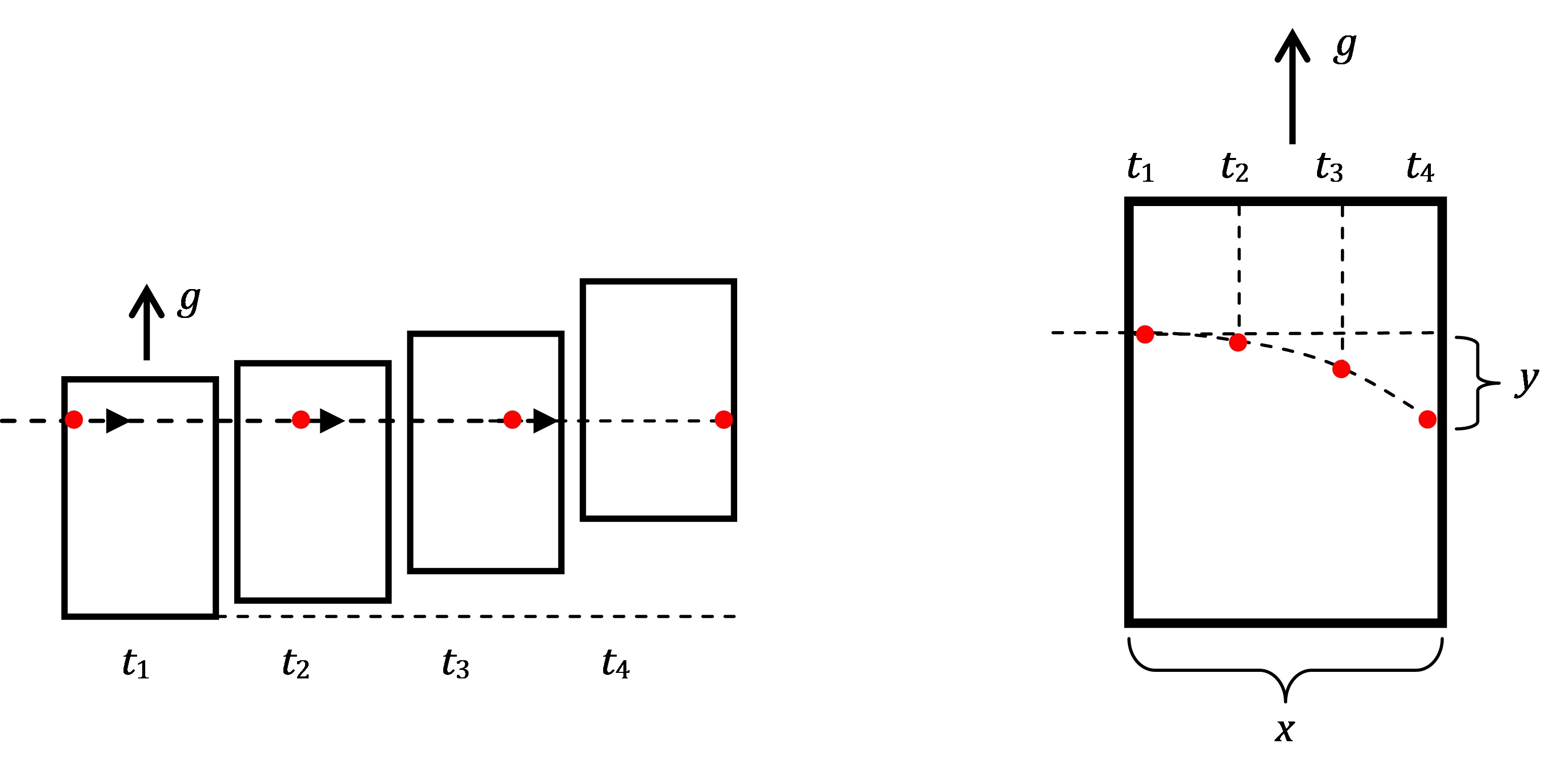}
  \caption{On the left, the trajectory of the photon is shown from outside the accelerated frame of reference, in four different instants of time. On the right, the trajectory of the photon is shown inside the accelerated frame of reference.}
\end{figure}

The external observer’s point of view is shown on the left image. This observer is in the same reference system from which the photon is emitted. To this observer, the photon follows a straight path. The point of view of an observer located inside the spaceship is shown on the right. The effect of the spaceship’s acceleration, as seen from inside, is a deviation of the photon’s path, which enters the ship through the upper left side and arrives at the lower right side. The photon travels a horizontal distance $x$, and deviates a vertical distance $y$. Thinking in terms of a beam of photons, i.e. a ray of light, the conclusion is the same. What is the angle of deviation of the photon in the ship’s frame of reference? To answer this question, we only need some considerations from kinematics.\\

The vertical distance that the photon travels is: 
 
\begin{equation}
y=\frac{1}{2} gt^{2}
\end{equation}

On the other hand, the time it takes the photon to fall this distance is given by:

\begin{equation}
t=\dfrac{x}{c}
\end{equation}

where $c = 3.00 \times 10^{8} m\cdot s^{-1}$ is the speed of light in the vacuum. Introducing this result into equation (1): 

\begin{equation}
y=\dfrac{x^{2}g}{2c^{2}}
\end{equation}

Due to the huge value of the speed of light, in general this deviation angle will be very small. Thus, if we call this angle $\alpha$, from equation (3) we can express it in radians ($rad$) as:

\begin{equation}
\alpha \approx \tan \alpha =\dfrac{y}{x} = \dfrac{xg}{2c^{2}}
\end{equation}

In virtue of the equivalency principle, we conclude that what happens to the ray of light in the frame of reference with constant acceleration $g$, will also occur in a uniform gravitational field $g$. That is to say, a ray of light inside a constant gravitational field $g$, will experience a deviation from its straight trajectory. This deviation is bigger as $g$ increases.\\

We can use the equivalency principle to approximate the deviation angle of a ray of light in a gravitational field of magnitude $g$. Let us consider a spherically symmetrical mass distribution, such a star. We will call this object lens because it deviates light rays that pass close to it. Behind the lens there is another star, which we call source. The source emits light that is deviated by the lens. Figure 3 illustrates this situation where the minimum distance between the light ray and the lens is $b$, and the angle of deviation is $\alpha$. In the figure, this angle is extremely magnified, because in reality it is very small, on the order of micro-seconds of arc. Due to this deviation of the light, a distant observer detects the source in an (apparent) position different from its real position.\\

\begin{figure}
  \centering
    \includegraphics[width=0.8\textwidth]{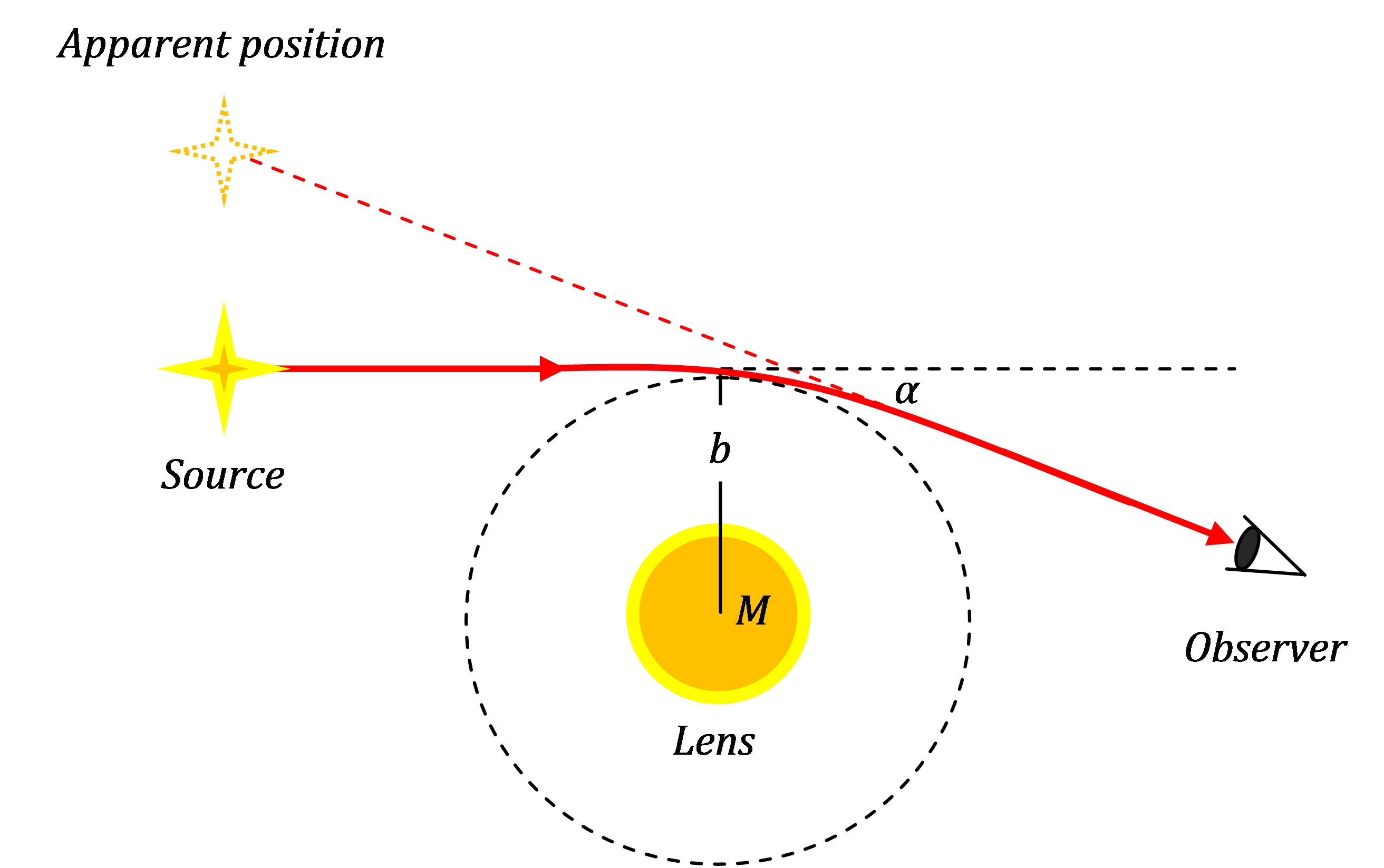}
  \caption{A ray of light (continuous red line) emitted from a source, is deviated by an angle $\alpha$ when passing close to a lens of mass $M$. The minimum distance between the light ray and the lens is $b$. The source is observed to be in a (apparent) position different from its real position. The region where the effects of the gravitational field generated by the lens are significant has a diameter $2b$.}
\end{figure}

Let $M$ be the mass of the lens, and let us suppose that $\sim b$ defines the diameter of the region where the effects of the gravitational field over the light beam are significant. If the lens is a common star (which excludes objects such as black holes or neutron stars), $g$ is weak, and we can use the law of universal gravitation. According to this law, the gravitational field’s magnitude of a gravitational lens at a distance $b$ from the ray of light is:

\begin{equation}
g=\dfrac{GM}{b^{2}}
\end{equation}

where $G = 6.67 \times 10^{-11} N\cdot m^{2}\cdot kg^{-2}$ is the universal gravitational constant. If we assume roughly that $g$ is given by equation (5) throughout the path of the light beam in the region of length $\sim2b$ (uniform gravitational field simplification), we can replace this value of $g$ in equation (4), taking $x = 2b$. Under these conditions we get:

\begin{equation}
\alpha \approx \dfrac{GM}{c^{2} b}
\end{equation}

Although this result is a very crude estimation, it only differs by a factor of 4 from the exact value found by Einstein in 1916 [1]:

\begin{equation}
\alpha = \dfrac{4GM}{c^{2} b}
\end{equation}

Let's imagine that the lens is our own Sun, which mass is $1.99 \times 10^{30} kg$, and let us assume that the beam of light passes very close to the Sun, such that $b$ is on the order of the solar radius: $6.95 \times 10^{8} m$. Because $2\pi rad = 360^{\circ}$, and each degree contains $3600^{''}$ (seconds), then $1 rad = (360\times 3600/2\pi)^{''}$, and therefore:  

\begin{equation}
\alpha = \dfrac{4(6.67\times 10^{-11} N \cdot m^{2} \cdot kg^{-2})(1.99 \times 10^{30}kg)}{(3\times 10^{8} m\cdot s^{-1})(6.95\times 10^{8} m)} \left(\dfrac{360\times3600}{2\pi}\right)^{''} = 1.75''
\end{equation}

This is the result predicted by Einstein in 1916 for the deviation of a ray of light in the vicinity of the Sun, and allowed the first observational confirmation of the theory of general relativity [1, 2]. As we shall see in the next section, equation (7) is the basis for the concept of Einstein ring.

\section{Gravitational lens and Einstein ring: Weighing a star with light} 

The gravitational lens effect is produced when the light originating from distant objects is deviated or curved around massive bodies, producing a phenomenon with significant similarities to what happens when light passes through a lens. The deviation of light near massive objects studied in the preceding section, is a particular case of the gravitational lens effect, where only a shift in the apparent position of a distant star is appreciated. In a more general situation, the effect can generate multiple images, increase in brightness and distortions in the object’s shape.\\ 

\begin{figure}
  \centering
    \includegraphics[width=0.8\textwidth]{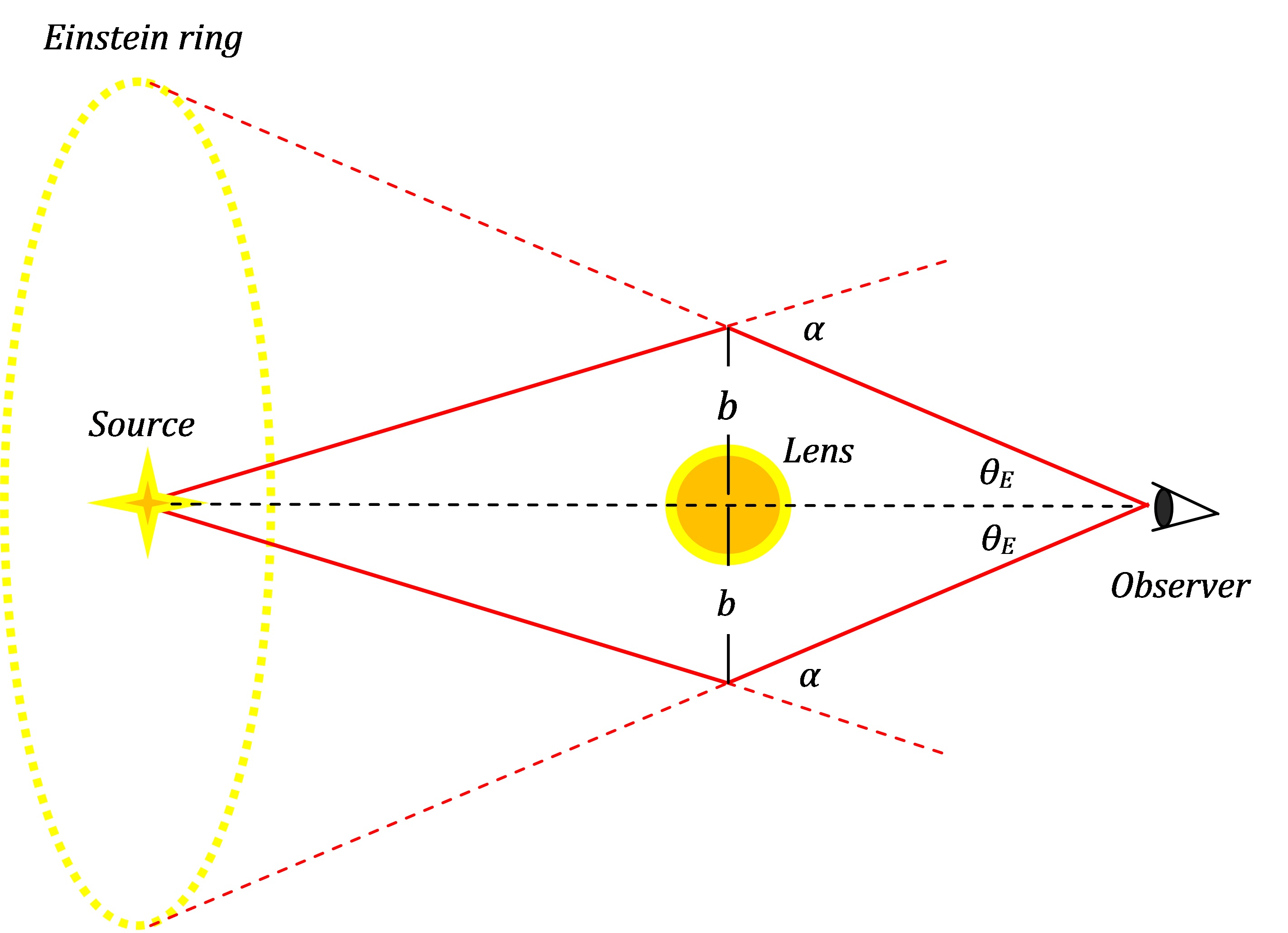}
  \caption{A source, a lens and an observer are perfectly aligned, which creates an Einstein ring. The drawing simplifies the phenomenon by showing only two of the light-rays that form the ring (continuous red lines), that subtend an Einstein’s angle $\theta_{E}$.}
\end{figure} 
 
The Einstein ring is the simplest and most symmetrical effect of gravitational lens, generated when the source, the lens and the observer are perfectly aligned. As its name suggests, the Einstein ring corresponds to the situation where the observer detects the light deviated by the lens as if it were a circumference or ring. Figure 4 illustrates this situation, where the Einstein ring is represented by a dashed yellow line circumference, centered on the source and being looked from the side. It can be observed that the rays of light that travel directly to the observer (continuous red lines) subtend an angle $\theta_{E}$, known as Einstein angle. Which is the characteristic magnitude of the Einstein ring. It also can be appreciated the deviation angle $\alpha$ calculated in the previous section. Again, this angle is very accentuated in the figure, because in practice, $\alpha$ and $\theta_{E}$ are on the order of microseconds of arc.\\
 
Figure 5 shows a version of figure 4 which is more useful to calculate $\theta_{E}$. Here are shown the distance between the source and the lens, $d_{LS}$, the distance between the lens and the observer, $d_{L}$, the distance between the source and the observer, $d_{S}$, and the angle $\beta$ between the undisturbed ray of light and the straight line between the source and the lens. To calculate $\theta_{E}$, let us note that by the exterior angle theorem we must have that: 
 
\begin{equation}
\alpha = \theta_{E} + \beta
 \end{equation} 
 
If $M$ is the mass of the lens, combining equations (7) and (9) we get:	

\begin{equation}
\alpha = \dfrac{4GM}{c^{2}b} = \theta_{E} + \beta
\end{equation}

Because $\alpha$ y $\theta_{E}$ are so small, when expressing them in radians we can use the approximations:

\begin{equation}
\theta_{E} \approx \tan \theta_{E} = \dfrac{b}{d_{L}},\hspace{1cm} \beta \approx \tan \beta = \dfrac{b}{d_{LS}}
\end{equation}

Introducing this expressions into equation (10):

\begin{equation}
\dfrac{4GM}{c^{2} b} = \dfrac{b}{d_{L}} + \dfrac{b}{d_{LS}}
\end{equation}

Solving for b:

\begin{equation}
b = \left(\dfrac{4GM}{c^{2}} \dfrac{d_{L} d_{LS}}{d_{L} + d_{LS}}\right)^{1/2} = \left(\dfrac{4GM}{c^{2}} \dfrac{d_{L}d_{LS}}{d_{S}}\right)^{1/2} 
\end{equation}

To express this result in $rad$, we divide member for member by the distance between the observer and the lens, obtaining Einstein angle:

\begin{equation}
\theta_{E} = \dfrac{b}{d_{L}} =  \left(\dfrac{4GM}{c^{2}} \dfrac{d_{LS}}{d_{L} d_{S}}\right)^{1/2} 
\end{equation}

For a more technical examination of this equation and the effect of gravitational lens, reference [11] is a publication that can be consulted. Let us note that $\theta_{E}$ increases as $M$ grows, as is to be expected, because an increase in mass implies a stronger gravitational field and a more pronounced deviation of the light rays close to the lens. From equation (5), taking $d_{LS} = d_{S} – d_{L}$ in equation (14) and solving for $M$, we get:

\begin{figure}
  \centering
    \includegraphics[width=0.7\textwidth]{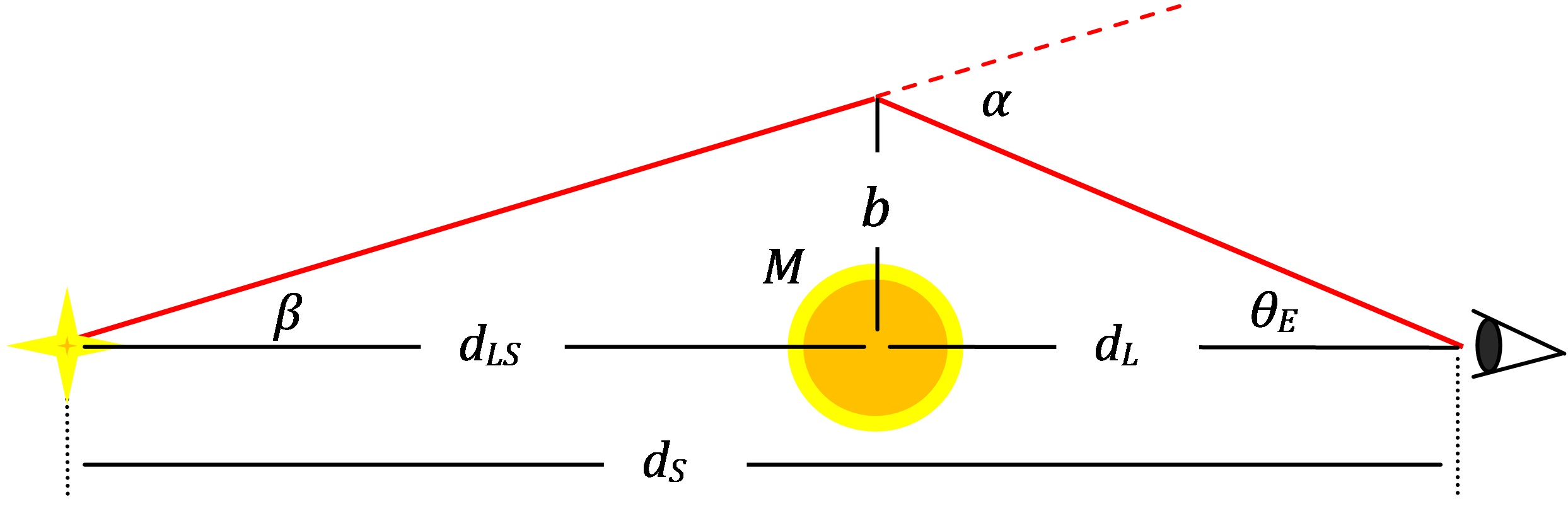}
  \caption{Reduced version of the figure 4, where we show Einstein angle, $\theta_{E}$, the distance between the source and the lens, $d_{LS}$, the distance between the lens and the observer, $d_{L}$, the distance between the source and the observer, $d_{S}$, and the angle $\beta$ between the undisturbed ray of light and the straight line between the source and the lens.}
\end{figure}

\begin{equation}
M=\dfrac{c^{2}}{4G} \dfrac{d_{L} d_{S} \theta_{E}^{2}}{(d_{S} - d_{L})} = \dfrac{c^{2}}{4G} \dfrac{d_{L} \theta_{E}^{2}}{(1- d_{L}/d_{S})}
\end{equation}

If we assume that $d_{L} \ll d_{S}$, which is the situation in which we are interested here, it is possible to neglect the term $d_{L}/d_{S}$, such that:

\begin{equation}
M=\dfrac{c^{2}}{4G} d_{L}\theta_{E}^{2}
\end{equation}

Obtaining the mass of a star from Einstein ring is a problem of considerable technical difficulties, among other reasons, because the probability of finding an astronomical system in a ring configuration is very low. However, it is possible to overcome this difficulty if a more probable scenario is considered, in which the source and the lens are slightly misaligned. In these conditions, an asymmetrical version of the Einstein ring is produced, which despite this, allows to obtain a good estimation of $\theta_{E}$ and the other parameters that define the ring. From these estimations, it’s possible to use equations (15) and (16) to calculate $M$ with very good accuracy. Leaving aside the complex technical details, this was the procedure to determine the mass of Stein 2051 B, that was done by a team of astronomers lead by Kailash Sabu from the Space Telescope Science Institute [4]. Using the Hubble Space Telescope, Sahu and his team tracked the white dwarf Stein 2051 B (lens) for two years, while it crossed in front of another star in the background (source). From the data acquired, the value obtained for the mass of Stein 2051 B was of $0.675 \pm 0.051 M_{\odot}$, where $M_{\odot} = 1.99 \times 10^{30} kg$ is the mass of the sun. \\

Let us reproduce the calculation done by Sahu's team, but leaving aside the technical issues and omitting the calculation of the errors. As reported in their article [4], the measured values are: $d_{L} = 5.52 parsec$ $(pc)$, $d_{S} = 2\times 10^{3} pc$ (note that $d_{L}/d_{S} \sim 10^{-3} \ll 1$), and $\theta_{E} = 31.53 milliarcsecond$ $(mas)$. To obtain the mass of the lens in kilograms ($kg$) we use that $1 pc = 3.09\times 10^{16} m$, and that $1 mas = (10^{-3})''$, where $1'' = (2\pi/360\times 3600) rad$. Introducing the corresponding values in equation (16):

\begin{equation}
M=\dfrac{(3\times 10^{8} m\cdot s^{-1})^{2}(17.06\times 10^{16} m)\left(\dfrac{63.06\pi}{1.296\times 10^{9}}rad \right)^{2}}{4(6.67 \times 10^{-11} N\cdot m^{2} \cdot kg^{-2})} = 1.34\times 10^{30}kg
\end{equation}

Using units of solar masses, it is found that $M = 0.67M_{\odot}$, which matches the value calculated by Sahu and his team.

\section{Final comments}

The gravitational lens effect constitutes a powerful astronomical research tool, which over the last decades has allowed the study of diverse phenomena. Depending on the goals, this type of research may focus on the source or on the gravitational lens. In the first case the studies rely on the fact that the lens can significantly intensify the light from the source, which has enabled the detection of very faint objects, such as far away galaxies and extrasolar planets with masses similar to Earth’s mass. In the second case, when studying the lens itself, the deviation of the light makes it possible to estimate the mass of the lens, which has allowed the study of big distributions of mass, such as galaxy clusters or the elusive dark matter.\\

Therefore, the usage of the gravitational lens effect to study the universe is not new, and is part of a wide research program developed long ago in different universities and astronomical centers throughout the world. However, what makes the work of Sahu and his team so relevant, is that they introduced a new and powerful procedure to calculate the mass of comparatively small objects that cannot be easily measured by other means (the gravitational lens effect applied to small objects, such as planets or stars, is known as microlensing). Furthermore, given the symmetry of the Einstein ring, the procedure can achieve higher accuracies than traditional methods.\\

Indeed, previous measurements of the mass of Stein 2051 B, based on its orbit around its companion star (recall that this white dwarf is part of a binary system) have underestimated its mass. This triggered a long controversy that appeared to put into question the validity of theoretical models of white dwarfs, based on the relation between mass and radius of these stars. The precise measurement of the mass of Stain 2051 B using Einstein ring, allowed to settle this controversy, supporting the validity of theoretical models. Hence, the achievement of Sahu and his team has had echoes that extend far beyond the measurement of the mass of a star. As this is a pioneer work, it is to be expected that in the following years new and important astronomical discoveries will be made based on Einstein ring.

\section{References}

[1] A. Einstein, Die Grundlage der allgemeinen Relativitätstheorie, Annalen der Physik, 354 (1916) 769-822.

\vspace{2mm}

[2] A. Einstein, The Collected Papers of Albert Einstein, Princeton University Press, Princeton, 1997.

\vspace{2mm}

[3] F.W. Dyson, A.S. Eddington, C. Davidson, A determination of the deflection of light by the Sun’s gravitational field, from observations made at the total eclipse of May 29, 1919 Philosophical Transactions of the Royal Society of London A, 220 (1920) 291-334.

\vspace{2mm}

[4] A. Einstein, Lens-like action of a star by the deviation of light in the gravitational field, Science, 84 (1936) 506-507.

\vspace{2mm}

[5] K.C. Sahu, J. Anderson, S. Casertano, H.E. Bond, P. Bergeron, E.P. Nelan, L. Pueyo, T.M. Brown, A. Bellini, Z.G. Levay, J. Sokol, M. Dominik, A. Calamida, N. Kains, M. Livio, Relativistic deflection of background starlight measures the mass of a nearby white dwarf star, Science, 356 (2017) 1046-1050.

\vspace{2mm}

[6] R.A. Serway, J.W. Jewett, Physics for Scientists and Engineers with Modern Physics, 7 ed., Thomson, Belmont, 2008.

\vspace{2mm}

[7] C.M. Becchi, M. D’Elia, Introduction to the Basic Concepts of Modern Physics, Springer, New York, 2010.

\vspace{2mm}

[8] C.H. Holbrow, J.N. Lloyd, J.C. Amato, E. Galvez, M.E. Parks, Modern Introductory Physics, 2 ed., Springer, New York, 2010.

\vspace{2mm}

[9] K. Krane, Modern Physics, 3 ed., John Wiley and Sons, Hoboken, 2012.

\vspace{2mm}

[10] P.A. Tipler, R.A. Llewellyn, Modern Physics, 6 ed., W. H. Freeman and Company, New York, 2012.

\vspace{2mm}

[11] A. Einstein, Über den Einfluß  der Schwerkraft auf die Ausbreitung des Lichtes, Annalen der Physik, 35 (1911) 898-908.

\vspace{2mm}

[12] B. Paczýnski, Gravitational microlensing in the local group, Annual Review of Astronomy and Astrophysics, 34 (1996) 419-459.

\end{document}